\documentclass[aps, prl, twocolumn, showpacs,
]{revtex4-1} 

\usepackage{amsmath}
\usepackage{amssymb}
\usepackage{graphicx}
\usepackage{color}

\renewcommand{\vec}[1]{\boldsymbol{#1}}
\begin{document}
	
\title{Towards Megajoule X-ray Lasers via  Relativistic 4-Photon Cascade in Plasma}
	\author{V.~M.~Malkin}
	\affiliation{Department of Astrophysical Sciences, Princeton University, Princeton, NJ USA 08540}
	\author{N.~J.~Fisch}
	\affiliation{Department of Astrophysical Sciences, Princeton University, Princeton, NJ USA 08540}

	\date{\today}

	\begin{abstract}
An extraordinarily efficient mechanism, operating at high laser intensities and powers, is identified for spectral transferring huge laser energies to shorter ultraviolet and x-ray wavelengths.  With megajoule laser energies currently available at near-optical wavelengths, this transfer would enable megajoule x-ray lasers, a huge advance over the millijoules x-ray pulses produced now.
	\end{abstract}
	\pacs{42.65.Sf, 42.65.Jx, 52.35.Mw}
	\maketitle

{ \sl Introduction:} 
The highest energy intense laser pulses are currently produced in the near-optical range through chirped pulse amplification mediated by material gratings \cite{Mourou98}. 
However, material gratings cannot be employed at much shorter wavelengths, so that available energies of intense laser pulses dramatically drop at ultraviolet and x-ray wavelengths.
In particular, the most energetic short x-ray pulses, currently produced by giant free-electron lasers,  are now only in the mJ range \cite{2016-RevModPhys-XFELphysics}. 
Hence, an efficient transfer of  laser energies from  near-optical wavelengths,  where megajoule laser energies are available \cite{2007-AplPhys-NIFlasers},  to deep ultraviolet and x-ray wavelengths 
would open up new research and technological frontiers.

The desired spectral energy transfer cannot be accomplished through  methods of high harmonic generation in gases   
\cite{1988-JourPhysB-Ferray-HighHarmonicInGases, 2012-Science-Popmintchev-XraysFromHHG}  or crystals  \cite{2011-NaturePhys-Ghimire-HighHarmInCrystal},   because gases and crystals cannot tolerate ultra-high intensities. 
Huge intensities can be tolerated in plasma.   
Several techniques do exploit the properties of plasma to generate very high intensity pulses through resonant 3-wave interactions, like  Raman backscattering  \cite{Malkin_99_PRL,Malkin_07_PRL,Ren_08_POP,Pai_08_PRL,Ping_09_POP,2017-SciRep-Vieux}, 
Brillouin scattering \cite{PRL-2013-Weber,2013-PoP-Riconda,2013-PoP-Lehmann-Spatschek,2017-PRE-Edwards-Fisch,2018-Nature-Kirkwood-BeamCombiner}, or magnetized low-frequency scattering \cite{2019-PRL-Edwards-Fisch}.
However, these techniques cannot be adapted to produce the significant frequency upshifts contemplated here. 

Huge frequency upshifts of laser energy could be achievable via resonant 4-photon cascade in plasma. In regimes where two photons are scattered into a higher frequency photon and a disposable photon of much lower frequency, each stage of the cascade would nearly double the photon frequency. 
Each stage would be useful in and of itself, by greatly increasing achievable laser energies at shorter, than previous stages, wavelengths.
Notably, cascading ten stages would dramatically convert  micrometer-wavelength laser energy to nanometer wavelengths.

\begin{figure}
	\centering
	\includegraphics[width=1.0\linewidth]{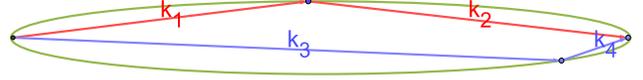}
	\caption{Resonant wavevector ellipsoid.}
	\label{f1}
\end{figure}

The classical synchronism conditions for the resonant 4-photon scattering in plasma are:
\begin{gather}
\label{e2} 
\vec{k}_1+\vec{k}_2=\vec{k}_3+\vec{k}_4\, ,\qquad
\omega_1+\omega_2=\omega_3+\omega_4\, .\,\\
\omega_j=\sqrt{k_j^2c^2+\omega_e^2}, \qquad \omega_e=\sqrt{4\pi n_0 e^2/m}.
\nonumber
\end{gather}
Here $c$ is the speed of light in vacuum; $m$ is the electron rest mass;  $-e$ is the electron charge; $n_0$ is the electron density; $\omega_e$ is the electron plasma frequency;  $ \omega_j$ are laser frequencies; and  $\vec{k}_j$ are  laser wavevectors in plasma. 
In a very undercritical plasma,  $\omega_e\ll \omega_j $, at moderately small,  $ k_j\gg k_{j\bot}\gg\omega_e/c $, transverse to $\vec{k}_{1}+\vec{k}_{2}=\vec{k}_{3}+\vec{k}_{4}$ components of all wavevectors, the frequency resonance condition reduces to $k_1+k_2\approx k_3+k_4$, like in vacuum, where the $\vec{k}$-vectors trace an ellipsoidal manifold, Fig.~\ref{f1}.

Intense laser pulses could be coupled in plasma through the relativistic electron nonlinearity. Most theoretical attention has been devoted to the most degenerate case of relativistic 4-photon coupling, in which all 4 photons are identical. The induced nonlinear frequency shift has important effects of relativistic self-modulation and self-focusing \cite{Litvak1969,Max1974,Sun1987,1993-PhysD-Malkin}, imposing a strict upper limit on powers $P$ of laser pulses propagating without filamentation,
\begin{equation}\label{e1}
P<P_{cr}={m^2c^5\omega^2}/{e^2\omega_e^2}\approx 
\, 17 \,({\omega}/{\omega_e})^2\,\rm GW . 
\end{equation}
For a petawatt laser power to be below $P_{cr}$, the laser-to-plasma frequency ratio $\omega/\omega_e$ needs to be so large, that the 4-photon coupling becomes too small to accomplish the spectral energy transfer within reasonable distances.

Because of the severity of this obstacle, and related obstacles, the high-power spectral transfer of huge laser energies in plasma appeared to be virtually impossible.  What we show here is how all the obstacles can in fact be overcome.  
To show this, we first need to get a sufficiently general 3-dimensional description of the relativistic 4-photon coupling in plasma, which has not yet been available in a form suitable for addressing this problem.

{ \sl Relativistic 4-Photon Coupling:} 
We begin with the Maxwell equations in Coulomb gauge and Hamilton-Jacobi equation for electron motion in electromagnetic fields  \cite{LandauLifshitz-ClassicalTheoryFields-1975-ch3} (kinetic effects are neglected, because  all the beating phase velocities and electron quiver velocities considered here are much larger than electron thermal velocities):  
\begin{gather}\label{e3}
\vec{H}=\nabla\times\vec{A}, \quad 
\vec{E}=- \partial_{ct}\vec{A}-\nabla\Phi,
\quad  \nabla\cdot\vec{A}=0, \\
\label{e4}
\Box\vec{A} \equiv
\partial^2_{ct}\vec{A} - \Delta\vec{A} = 4\pi\vec{J}/c  - \partial_{ct}\nabla\Phi,\\
\label{e5} 
\vec{J}=-en\vec{P}/\sqrt{m^2+P^2/c^2}, \quad 
\Delta\Phi = 4\pi e (n-n_0), \\
\label{e6} 
\!\!\!\!\!\vec{P}=e\vec{A}/c+\nabla S, \;\; 
 \partial_{t} S = e\Phi - c\sqrt{m^2c^2+P^2}+mc^2.\!
\end{gather}
By introducing dimensionless electromagnetic potentials and electron momentum, and rescaling the action function $S$ by the factor $mc^2$,
\begin{equation}\label{e7}
\!\vec{A} = mc^2\vec{a}/e, \;\; \Phi=mc^2\phi /e, \;\;
\vec{P} = mc\vec{p}, \;\; S=mc^2 s,\!
\end{equation}
the equations can be presented in the form
\begin{gather}
\label{e8}
c^2\Box\vec{a} = -\vec{p} (1+p^2)^{-1/2}(\omega_e^2+c^2\Delta\phi)  - c\partial_{t}\nabla\phi,\\
\label{e9} 
\nabla\cdot\vec{a}=0, \;\;
\vec{p}=\vec{a}+c\nabla s, \;\; 
\partial_{t} s = \phi - \sqrt{1+p^2}+1.
\end{gather}
For mildly relativistic electron quiver velocities in laser pulses, $a\ll 1$, Eqs.~(\ref{e8})-(\ref{e9}) can be expanded in powers of the small parameter $a$. To calculate the 4-photon coupling, the expansion should include terms up to cubic in $a$. 
With laser beatings well off the plasma wave resonance, the leading term of the electrostatic potential $\phi$ expansion is quadratic in $a$, 
so that the cubic in $a$ expansion of Eq.~(\ref{e8}) is
\begin{equation}\label{e10}
 (c^2\Box + \omega_e^2)\vec{a} = \vec{a}(a^2\omega_e^2/2 - c^2\Delta\phi) - c(\omega_e^2 \nabla s + \nabla\partial_{t}\phi).
\end{equation}
Equation $\nabla\cdot\vec{a}=0$ can be used to exclude $s$ from Eq.~(\ref{e10}).
In a uniform plasma, $\nabla\omega_e=0$, it gives 
\begin{gather}\label{e11}
\omega_e^2 s +\partial_{t}\phi=\Delta^{-1}\nabla\cdot[\vec{a}(a^2\omega_e^2/2- c^2\Delta\phi)]/c, \\
\label{e12}
\!\!\!\!\!\! (c^2\Box + \omega_e^2)\vec{a} = (1-\nabla\Delta^{-1}\nabla\cdot)
[\vec{a}(a^2\omega_e^2/2 - c^2\Delta\phi)]/c. \!\!
\end{gather}
Eqs.~(\ref{e11}) and $\partial_{t} s = \phi - \sqrt{1+p^2}+1 $ can be used now to  exclude $\phi$ from Eq.~(\ref{e12}),
\begin{gather}\label{e13}
(\partial^2_{t} + \omega_e^2)\phi= a^2\omega_e^2/2,\\
\!\!\!\!\!\!\!\!(c^2\Box\! + \!\omega_e^2)\vec{a}\!\! =\!\!(1\! -\!\nabla\!\Delta^{\!\! -\! 1}\nabla\cdot) 
\vec{a}[1\! -\! c^2\!\Delta(\partial^2_{t}\! + \!\omega_e^2)^{\! -\! 1} ] a^2\!\omega_e^2/2.\!\!
\label{e14}\end{gather}
For paraxial laser pulses of several different frequencies, the beatings of $a^2$, at frequencies that well exceed $\sqrt{\omega_e^2+k_\bot^2 c^2}$, propagate with speeds close to $c$. 
For such beatings, the second term in the square brackets Eq.~(\ref{e14}), associated with the electrostatic potential $\phi$, nearly exactly compensates the first term in the square brackets, associated with the relativistic variation of electron mass. 
The compensation significantly reduces the 4-photon coupling for paraxial laser pulses. 
This important effect is missed in calculations of the 4-photon scattering probability \cite{1979-ApplPhysLett-Praddaude-Cohrnt4WaveScatPlasm, 1991-IEEETransPlasmSci-Federici-4WMixingPlasm} neglecting the relativistic variation of the electron mass.

Four resonant laser pulses, apart from their resonant interaction, produce small non-resonant beatings $\delta\vec{a}$,
\begin{gather}\label{e15}
\vec{a}=\sum_{j=1, \sigma=\pm}^{j=4} \vec{a}_{\sigma j}\exp[\imath (\vec{k}_{\sigma j}\vec{r}-\omega_{\sigma j} t)] +\delta\vec{a}, \\
\vec{k}_j\cdot \vec{a}_j=0,\;\;\vec{a}_{-j}=\vec{a}_j^\ast,\;\; \vec{k}_{-j}=-\vec{k}_{j}, \;\; 
\omega_{-j}=-\omega_{j} .\label{e16}
\end{gather}
Substituting Eq.~(\ref{e15}) into Eq.~(\ref{e14}) and collecting the resonant terms  leads to the following equations for slowly varying canonical amplitudes \cite{1997-UFN-Zakharov-HamiltonianFormalismNonlWaves} $b_j=a_j\sqrt{\omega_j}\,$:
\begin{gather}
\left[\imath\left(\frac{\partial}{\partial t} + \frac{c^2\vec{k}_j\cdot\nabla}{\omega_j} \right) - \frac{c^2\Delta_{j\bot}}{2\omega_j}  \right] b_j =\delta\omega_j b_j + \frac{\partial\mathcal{H}}{\partial b_j^\ast}, \label{e17} \\
\mathcal{H}=Vb_1b_2b_3^\ast b_4^\ast +c.c.,  \label{e18} \\
\!\! \delta\omega_j=\frac{\omega_e^2}{2\omega_j}\left(  \sum_{l=1,\sigma=\pm}^{l=4,\sigma l\neq j}\frac{|b_l  e_{j,-\sigma l}
 	|^2}{\omega_l} f_{j,-\sigma l}
 -\sum_{l=1}^{l=4}\frac{|b_l|^2}{\omega_l} \right), \label{e19}\\
 \nonumber
\! V\!=\!\frac{\omega_e^2 (f_{1,2}e_{1,2}e_{-3,-4}\! +\! f_{1,-4}e_{1,-4}e_{2,-3}\!  +\! f_{2,-4}e_{1,-3}e_{2,-4})} {2\sqrt{\omega_1\omega_2\omega_3\omega_4}},\\
 e_{j,l}=\frac{\vec{a}_j\cdot\vec{a}_l}{a_ja_l},\;\;\;\;
f_{j,l}=\frac{c^2(\vec{k}_j+\vec{k}_{l})^2}{(\omega_j+\omega_{l})^2-\omega_e^2}-1
\label{e20} 
\end{gather}
($\Delta_{j\bot}$ is the Laplacian in the plane perpendicular to $\vec{k}_j$). 

{ \sl Transverse Filamentation Instability:}
For a single laser pulse, $b_1=a\sqrt{\omega},\; b_2=b_3=b_4=0$, Eqs.~(\ref{e17})-(\ref{e20})  reduce to the standard cubic nonlinear Schroedinger equation,
\begin{equation}\label{e21}
\left[\imath\left(\frac{\partial}{\partial t} + \frac{c^2 k}{\omega} \frac{\partial}{\partial z} \right) - \frac{c^2\Delta_\bot+\omega_e^2|a|^2}{2\omega }  \right] a = 0.
\end{equation}
A spatially uniform solution of this equation, 
\begin{equation}\label{e22}
a=a_0\exp(-\imath \Gamma t), \quad \Gamma=|a_0|^2\omega_e^2/2\omega,
\end{equation}
can experience  small transverse modulations
\begin{gather}
\nonumber \!\tilde{a}\!=\!
\exp(-\imath \Gamma t)\{\psi\exp[\imath(\vec{\kappa}\vec{r\!}_\bot\!\! -\!\Omega t)]\! + \!\chi^\ast\!\exp[-\imath(\vec{\kappa}\vec{r\!}_\bot\!\! -\!\Omega^\ast t)]\} ,\\
\label{e23}
\Omega^2=c^2\kappa^2/2\omega(c^2\kappa^2/2\omega-2\Gamma)\equiv \Omega^2_\kappa\, .
\end{gather}
At $\kappa < \sqrt{2}\, a_0\omega_e/c$, the modulations are unstable. The largest growth rate, reached at $\kappa=a_0\omega_e/c$, is $\Gamma$.  Pulses of powers exceeding the critical power Eq.~(\ref{e1}) have apertures sufficient to accommodate unstable modulations.

For two laser pulses, such that $|b_2|\approx |b_1|$, $b_3=b_4=0$,  $\vec{e}_{1,-2}\approx 1$,  $|\vec{k}_2-\vec{k}_1|\ll k_1$, and 
$\omega_1\gg \omega_2-\omega_1 > \omega_e$, Eqs.~(\ref{e17})-(\ref{e20}) reduce to
\begin{gather}
\left[\!\imath\left(\!\frac{\partial}{\partial t}\! +\! \frac{c^2\vec{k}_1\!\cdot\!\nabla}{\omega_1} \right)\! + \! \frac{\omega_e^2(G|a_2|^2\! -\! |a_1|^2 )\! -\! c^2\Delta_{1\bot}}{2\omega_1 }  \right] a_1 = 0,  \nonumber\\
\left[\!\imath\left(\!\frac{\partial}{\partial t}\! +\! \frac{c^2\vec{k}_2\!\cdot\!\nabla}{\omega_2} \right)\! +\! \frac{\omega_e^2(G|a_1|^2\! -\!|a_2|^2 )\! -\! c^2\Delta_{2\bot}}{2\omega_2 }  \right] a_2 = 0,  \nonumber\\
G\approx f_{1,-2} -1. 
  \label{e24}
\end{gather}
These equations have a spatially uniform solution 
\begin{equation}\label{e25}
a_2=a_1=a_0\exp[\imath (G-1)\Gamma t] 
\end{equation}
with $\Gamma $ of Eq.~(\ref{e22}). Small modulations of this solution in the direction transverse to $\vec{k}_1+\vec{k}_2 $ satisfy the dispersion equation 
\begin{equation}\label{e26}
\!\!\left[\!\!\left(\frac{\Omega}{\kappa c}\!-\!\frac{{k}_\bot c}{\omega}\right)^{\!\! 2}\!\! - \! \frac{\Omega^2_\kappa}{\kappa^2c^2}\right]\!\! \left[\!\!\left(\frac{\Omega}{\kappa c}\! +\!\frac{{k}_\bot c}{\omega}\right)^{\!\! 2}\!\! - \!\frac{\Omega^2_\kappa}{\kappa^2c^2}\right]\! = \!
\frac{G^2\Gamma^2}{\omega^2},\!\!
\end{equation} 
where $ \vec{k}_\bot \equiv\vec{k}_{1\bot}=-\vec{k}_{2\bot}$ and $\Omega^2_\kappa$ is given by Eq.~(\ref{e23}).
As seen from Eq.~(\ref{e26}), modulations of pulses 1 and 2 are weakly coupled for 
$k_\bot^2\gg G\Gamma\omega/c^2$, a condition  satisfied for $(\omega_2-\omega_1-\omega_e)k_\bot^2c^2/\omega_e^2\gg |a_0|^2\omega_e$.
In the opposite limit, the modulations are strongly coupled and $\Omega^2\approx \Omega^2_\kappa \pm G\Gamma\kappa^2c^2/\omega$. 
The ``$+$" branch is stabilized for $G> 1$, but then the ``$-$" branch gets even more unstable than for a single pulse. 
Thus, achieving the defocussing nonlinear frequency shift in the uniform solution (\ref{e25})  for $G>1$ does not prevent the transverse filamentation instability.  
Modulations of collinear laser pulses are necessarily strongly coupled  and unstable at total powers exceeding the critical power. (This explains why achieving a defocussing nonlinear frequency shift for collinear laser pulses, \cite{2006-PoP-Shukla-2NonlCoupledLaserPlasma, 2008-PRE-Shvets-OpticalSuppresRelatvstcSelfFoc}, did not help increase to any significant degree the laser power propagating in plasma without filamentation, as was borne out by numerical simulations \cite{2008-PRE-Shvets-OpticalSuppresRelatvstcSelfFoc}.) 
In contrast, modulations of non-collinear laser pulses can easily satisfy conditions for the weakly coupled regime, and be stable to transverse modulations at each pulse power not exceeding its own critical power.  
The total power of all pulses  can then significantly exceed the individual critical powers.

{ \sl Linear Growth Rate of 4-Photon Amplification:}
Resonant 4-photon amplification can be initiated by two pump pulses 1 and 2, and small seed pulse 3. In the paraxial geometry with the axis along $\vec{k}_{1}+\vec{k}_{2}=\vec{k}_{3}+\vec{k}_{4}$, so that $\vec{k}_{2\bot}=-\vec{k}_{1\bot}$, $\vec{k}_{4\bot}=-\vec{k}_{3\bot}$, the frequency resonance condition reduces to ${k_{1\bot}^2}{k_3 k_4}\approx {k_{3\bot}^2}{k_1 k_2} $.
For moderately close pump frequencies, $\omega_1\gg \omega_2-\omega_1\gg \omega_e $, and moderately large ratio of seed frequencies, 
$\omega_3\gg \omega_4$, the pump energy mostly goes into the amplified pulse 3, whose frequency is nearly twice the pump frequencies. 
When all polarizations are the same, the 4-photon coupling coefficient Eq.~(\ref{e20}) reduces to 
\begin{equation}\label{e30}
V\approx {3\,\omega_e^2 \, k_{1\bot}^2} (2\omega_1\sqrt{\omega_3\omega_4}\, k_1^2)^{-1}.
\end{equation} 
The spatially uniform solution of Eqs.~(\ref{e17})-(\ref{e20}) has then small seeds 3 and 4 growing exponentially with the rate
\begin{equation}\label{e31}
\gamma= V|b_1 b_2|\approx {3\,\omega_e^2 \, k_{1\bot}^2|a_1|^2} (2\sqrt{\omega_3\omega_4}\, k_1^2)^{-1}.
\end{equation}
For example, for $a_1=0.1$, $k_{1\bot}= k_1/7$, $\omega_4=\omega_1/5$, the rate is $\gamma\approx 4.7\times 10^{-4}\omega_e^2/\omega_1$, corresponding to the pulse amplification distance $c/\gamma\approx 3.3\times 10^2 \lambda_1 \omega_1^2/\omega_e^2$. 
For  $\omega_1=50\, \omega_e$, it is $c/\gamma\approx 8.4\times 10^5 \lambda_1$. 
For the laser wavelength $\lambda_1\approx 350\,$nm, as at NIF   \cite{2007-AplPhys-NIFlasers}, $c/\gamma\approx 30\,$cm. At succeeding steps of the spectral energy transfer, occurring at shorter laser wavelengths, even shorter distances would suffice.

{ \sl Control of Nonlinear Detuning:}
As the seed amplitude and the pump depletion grow, the initially perfect resonance may be detuned by  nonlinear frequency shifts $\delta\omega_j$. Notably, these shifts, unlike the 4-photon coupling, do not exhibit automatic cancellation of the leading terms for paraxial pulses and, thus, can much exceed the coupling.
The detuning, $\delta\omega=\delta\omega_4+\delta\omega_3-\delta\omega_2 -\delta\omega_1$,
can be controlled by using ``dual seeds", coupled like pump pulses 1 and 2. 
Let all pulses have the same polarization, and let seed pulse 5 be close to 3 in amplitude $|b_5|\approx|b_3|$,  and frequency $\omega_5\gg \omega_3-\omega_5\gg \omega_e $, while $\vec{k}_{5\bot}\approx - \vec{k}_{3\bot} $. 
Pulse 6, resonantly amplified with  pulse 4 by the same pumps 1 and 2,  would then satisfy conditions  $\omega_4\gg \omega_6-\omega_4\gg \omega_e $,  $\vec{k}_{6\bot}\approx - \vec{k}_{4\bot} $, $|b_6|\approx|b_4|\approx|b_3|$. 
The nonlinear detuning of each of the dual resonances would then be
\begin{gather}
\!\!\delta\omega\!\approx\! \frac{\omega_e^2 |b_4|^2}{2\omega_4^2} \!\left[\!\frac{(\vec{k}_ 6\!\!-\!\!\vec{k}_4)^2}{(k_6\!\! -\!\! k_4)^2}\!\! -\!\! 3\!\! -\!\!\frac{2\omega_4}{\omega_3}\!\! +\!\!\frac{4\omega_4}{\omega_1}\!\right]\!\! +\!\!
\frac{\omega_e^2 |b_3|^2}{2\omega_3^2}\! \left[\!\frac{(\vec{k}_3\!\! -\!\!\vec{k}_5)^2}{(k_3\!\! -\!\! k_5)^2}\!\! 
\right.\nonumber\\ 
\left.  -\! 3\! - \!\!\frac{2\omega_3}{\omega_4}\!\! +\!\!\frac{4\omega_3}{\omega_1}\!\right]\!-\!
\frac{\omega_e^2 |b_1|^2}{\omega_1^2}\! \left[\!\frac{(\vec{k}_2\!\! -\!\!\vec{k}_1)^2}{(k_2\!\! -\!\! k_1)^2}\!\! -\! 3\! +\! \frac{\omega_1}{\omega_4}\! +\!\frac{\omega_1}{\omega_3}\!\right]. \label{e32}
\end{gather} 
It can be zeroed out by proper selection of the ratios between the transverse and longitudinal components of vectors $\vec{k}_6-\vec{k}_4=\vec{k}_3-\vec{k}_5$ and $\vec{k}_2-\vec{k}_1$.

{\sl Pump Depletion for Fully Overlapping Pulses:}
At zeroed out resonance detuning, there is a simple analytical solution of Eqs.~(\ref{e17})-(\ref{e20}), where all $b_j$ keep constant phases synchronized such that $\arg b_6+\arg b_5=\arg b_4+\arg b_3=\arg b_2+\arg b_1-\pi/2$, while the intensities are given by the formulas
\begin{gather}\nonumber
|b_3|^2=|b_4|^2=|b_5|^2=|b_6|^2\equiv I_{seed}/4, \\
|b_1|^2=|b_2|^2\equiv I_{pump}/2, \;\; I_{seed}=I_{seed0}+I_{pump0}-I_{pump}\nonumber\\
I_{pump}=
\frac{(I_{pump0}+I_{seed0})I_{pump0}}{I_{pump0}+I_{seed0} \exp(2\gamma t)}\, .\label{e33}
\end{gather}
Within several growth times $\gamma^{-1}$, nearly all the pump energy is converted into the seed pulses 3 and 5 of nearly doubled frequency. Eq.~(\ref{e33}) can be generalized for multiple pumps amplifying the same seeds.
For the spatially uniform solution to be applicable, all pulses should nearly completely overlap throughout the amplification process.

{\sl Longitudinal Pulse Slippage Regime:}
There are also regimes affected by the pulse slippage, or even usefully employing it. For example, let a small seed pulse 3 enter at $t=0$ the rear edge of fully overlapping pump pulses 1 and 2 having close frequencies and equal numbers of photons. 
The 4-photon amplification occurs at $t>0$ in the domain $c_1 t< z < c_3 t\leq c_1 t+L$, where $c_j \approx c (1-k_{j\bot}^2/2k_j^2)$ is the longitudinal group velocity of pulse $j$, and $L$ is the pump length. 
At an advanced nonlinear stage, the amplified pulse amplitude is 
\begin{gather}\nonumber
|b_3|\approx |b_{10}|\sqrt{k_4/k_3}\, \{1+\exp[2({\zeta_M}-\zeta)]\}^{-1/2} \tau/\zeta_M,\\ 
\tau=2\gamma t,\;\; \zeta= 2\gamma\sqrt{t(c_3t-z)/(c-c_4)},\label{31}\\
\zeta_M\exp\zeta_M \approx \tau \sqrt{k_4/k_3}\, |b_{10}|/|b_{30}|\gg 1 \label{32}
\end{gather}
in the domain $\zeta > 0, \; \zeta -\zeta_M\ll \zeta_M$; in the overlapping domain $\zeta -\zeta_M\gg 1$, where pumps are already depleted,
\begin{equation}\label{33}
\!|b_3|\!\approx\! |b_{10}|\sqrt{k_4/k_3}\, \tau\tilde{\zeta}_M/\zeta^2, \;
\tilde{\zeta}_M\!\approx \!\zeta_M\! +\!\ln (\zeta_M \tilde{\zeta}_M/\zeta^2) .\!\!\!\!
\end{equation}
The pulse amplitude and energy (located at $\zeta\sim \zeta_M$) grow approximately linearly with the amplification time, while the pulse duration decreases approximately inversely to the amplitude.

{ \sl Discussion:} 
We show how laser powers far exceeding critical could be safely handled in multi-pulse weakly-coupled dynamic regimes, thus overcoming the major limitation of strongly-coupled regimes, 
by choosing pulses insufficiently collinear to act synchronously in triggering filamentation. As seen from Fig. 1, many different pairs of insufficiently collinear pump pulses can be in 4-photon resonance with the same amplified seed simultaneously. Similarly, multiple, insufficiently collinear seeds could be simultaneously amplified by the same set of pump pulses. Thus, each pump and each seed power can stay below the critical power, while the combined power can be much larger than critical. 

We show how to stay within the 4-photon resonance throughout the entire amplification process, despite
varying nonlinear frequency shifts much exceeding the resonance bandwidth, 
by using novel ``dual seeds" securing mutual cancellation of the frequency shifts in the synchronism conditions. An incidental benefit is that unwanted resonances may not survive.

Keeping exact resonance  does not by itself guarantee a highly efficient energy transfer from pumps to pumped pulses, since the energy flow could reverse back to pumps, if a pump amplitude is allowed to go through zero. Even if the process stopped, say, by reaching the edge of plasma, the remaining pump energy stays unused. We show how to avoid these losses, by using pumps with the same number of photons. Such pumps are depleted simultaneously, which ensures nearly total energy transfer.  
Any small leftover of pump energy, due to inexact matching of numbers of photons, just slightly reduces the efficiency. The process can realistically be terminated for many pump pulses simultaneously, before any reversion of the energy flow occurs, because the rate of energy transfer drops significantly when pump amplitudes become small. For example, if the pump leftover energy is 10\% of the initial pump energy, the distance within which pumps stay that small is 10 times larger than the initial pump depletion distance, which leaves an ample space for proper stopping the process.

The means used here for preventing the reverse energy flow might suggest more general tools for manipulating inverse cascades, since the reverse energy flow from seeds to pumps, prevented here, is the dynamic counterpart of kinetic inverse energy cascades. In particular, this analogy might have important implications for improving
the kinetic method of suppressing the relativistic filamentation instability by phase randomization of powerful laser pulses \cite{2016-PRL-Malkin,2018-PRE-Malkin-OpticalTurb}. That method is seriously impeded  by the well-known tendency of Bose-Einstein condensate formation via the inverse energy cascade \cite{1992-ZakharovLvovFalkovich-KolmogorovTurb, 1996-PRL-Malkin, 2015-PRE-Falkovich}. 
 
The effects we employ to prevent the filamentation are obviously 3-dimensional, therefore transverse slippage of the pulses could be an issue. The slippage, however, can be neglected for large aperture pulses, corresponding to the large powers that we consider here.  For example, 30 cm aperture paraxial pulses, like those used at National Ignition Facility (NIF) \cite{2007-AplPhys-NIFlasers}, can easily be kept nearly fully overlapped within 30 cm interaction length. For pulses fully overlapped in the transverse directions, we presented simple analytical solutions of the evolution equations, showing energy transfer to a nearly double-frequency seed up to the total pump depletion, addressing both the cases of negligible or substantial longitudinal slippages.

Although we do not emphasize here applications, it  can be expected that short-wavelength MJ laser pulses will enable radically new discoveries and technologies. 
An example of  new technologies that might be readily imagined is the delivery of laser power to the compressed target core for achieving fast ignition in inertial confinement fusion, not with lasers in the optical range \cite{tabak}, but with even kilojoules at x-ray wavelengths capable of naturally penetrating even very dense plasma layers.  

Notably, in our scheme, plasma need not be very homogeneous. We propose to use a plasma of very undercritical density, where the electron plasma frequency is much smaller than the laser frequencies, so that the 4-photon resonance synchronism conditions are basically the same as in vacuum. We identify regimes where such a rarefied plasma provides the 4-photon coupling sufficient for accomplishing the spectral energy transfer to shorter wavelengths within reasonable distances. In these regimes, plasma inhomogeneities do not spoil the useful resonance.  At the same time, the very same plasma inhomogeneities may, in fact, serve to suppress parasitic processes such as Raman or Brillouin scattering, mediated by plasma waves or sound waves, which are much more sensitive to inhomogeneities than the photons that we consider \cite{malkin2000prl,solodov2003pop}. 
The parasitic processes involving plasma waves can be further suppressed by laser frequency chirping \cite{Toroker_12_PRL}, without affecting the useful 4-photon resonance.

{\sl Summary:} 
The proposal advanced here, to transfer megajoule energies of short laser pulses to deep-ultraviolet and x-ray wavelengths, is unique in its ambition.
It features also a concise general description of  resonant relativistic 4-photon scattering of laser pulses, filling an important gap in the theory of basic nonlinear interactions in plasma.
It identifies the methodologies that can be used for a highly efficient plasma-based spectral transfer of huge laser energies to shorter ultraviolet and x-ray wavelengths. Apart from the evident importance of this application, the methodologies put forth upon which it relies are also of basic academic significance.

{ \sl Acknowledgment:}  
This work was supported by NNSA DE-NA0003871 and AFOSR FA9550-15-1-0391.


\providecommand{\noopsort}[1]{}\providecommand{\singleletter}[1]{#1}%

\end{document}